\documentclass[12pt]{article}
\usepackage{epsfig}

\def\Rad#1{%
	\begingroup
	\def\RadTempCs{{#1}}\let\RdxTempCs=\empty
}

\def\DoRad{%
	\relax
	\ifx \RdxTempCs\empty
		\sqrt\RadTempCs
	\else
		\root \RdxTempCs \of \RadTempCs
	\fi
	\endgroup
}

\begin{document}

\title{Power Markets for Controlling Smart Matter}
\author{Oliver Guenther, Tad Hogg and Bernardo A. Huberman \\
	Xerox Palo Alto Research Center \\
	Palo Alto, CA 94304}

\maketitle

\begin{abstract}
Embedding microscopic sensors, computers and actuators
into materials allows physical systems to actively monitor and respond
to their environments. This leads to the possibility of creating smart
matter, i.e., materials whose properties can be changed under program
control to suit varying constraints. A key difficulty in realizing the
potential of smart matter is developing the appropriate control
programs. We present a market{--}based multiagent solution to the
problem of maintaining a physical system near an unstable configuration,
a particularly challenging application for smart matter. This market
control leads to stability by focussing control forces in those parts of
the system where they are most needed. Moreover, it does so even when
some actuators fail to work and without requiring the agents to have a
detailed model of the physical system.
\end{abstract}

\section{ Introduction}
Embedding microscopic sensors, computers and actuators into
materials allows physical systems to actively monitor and respond to
their environments in precisely controlled ways. This is particularly so
for microelectromechanical systems (MEMS)~\cite{berlin95,bryzek94,web.mems96} 
where the devices are
fabricated together in single silicon wafers. Applications include
environmental monitors, drag reduction in fluid flow, compact data
storage and improved material properties. 

In many such applications the relevant mechanical processes are
slow compared to sensor, computation and communication speeds. This
gives a {\em smart matter} regime, where control programs
execute many steps within the time available for responding to
mechanical changes. A key difficulty in realizing smart matter{'}s
potential is developing the control programs. This is due to the need to
robustly coordinate a physically distributed real-time response with
many elements in the face of failures, delays, an unpredictable
environment and a limited ability to accurately model the system{'}s
behavior. This is especially true in the mass production of smart
materials where manufacturing tolerances and occasional defects will
cause the physical system to differ somewhat from its nominal
specification. These characteristics limit the effectiveness of
conventional control algorithms, which rely on a single global processor
with rapid access to the full state of the system and detailed knowledge
of its behavior. 

A more robust approach for such systems uses a collection of
autonomous agents, that each deal with a limited part of the overall
control problem. Individual agents can be associated with each sensor or
actuator in the material, or with various aggregations of these devices,
to provide a mapping between agents and physical location. This leads to
a community of computational agents which, in their interactions,
strategies, and competition for resources, resemble natural
ecosystems~\cite{Huberman88Eco}. Distributed
controls allow the system as a whole to adapt to changes in the
environment or disturbances to individual components~\cite{hogg91a}.

Multiagent systems have been extensively studied in the context of
distributed problem solving~\cite{durfee91,gasser89,lesser95}. They have also been
applied to problems involved in acting in the physical world, such as
distributed traffic control~\cite{nagel94},
flexible manufacturing~\cite{upton92}, the design
of robotic systems~\cite{sanderson83,williams96a},
and self-assembly of structures~\cite{semela95}.
However, the use of multiagent systems for controlling smart matter is a
challenging new application due to the very tight coupling between the
computational agents and their embedding in physical space.
Specifically, in addition to computational interactions between agents
from the exchange of information, there are mechanical interactions
whose strength decreases with the physical distance between them.

In this paper we present a novel control strategy for unstable
dynamical systems based on market mechanisms. This is a particularly
challenging problem, for in the absence of controls, the physics of an
unstable system will drive it rapidly away from the desired
configuration. This is the case, for example, for a structural beam
whose load is large enough to cause it to buckle and break. In such
cases, weak control forces, if applied properly, can counter departures
from the unstable configuration while they are still small. Successful
control leads to a virtual strengthening and stiffening of the material.
Intentionally removing this control also allows for very rapid changes
of the system into other desired configurations. Thus an effective way
of controlling unstable systems opens up novel possibilities for making
structures extremely adaptive.

\section{ Dynamics of Unstable Smart Matter}
The devices embedded in smart matter are associated with
computational agents that use the sensor information to determine
appropriate actuator forces. The overall system dynamics is a
combination of the behavior at the location of these agents and the
behavior of the material between the agent locations. In mechanical
systems, displacements associated with short length scales involve
relatively large restoring forces, high frequency oscillations and rapid
damping. Hence, they are not important for the overall
stability~\cite{hogg96b}. Instead, stability is
primarily determined by the lowest frequency modes. We assume that there
are enough agents so that their typical spacing is much smaller than the
wavelengths associated with these lowest modes. Hence, the lower
frequency dynamics is sufficiently characterized by the displacements at
the locations of the agents only. The high-frequency dynamics of the
physical substrate between agents serves only to couple the agents{'}
displacements.

\begin{figure}
\hspace*{\fill}
\epsfig{file=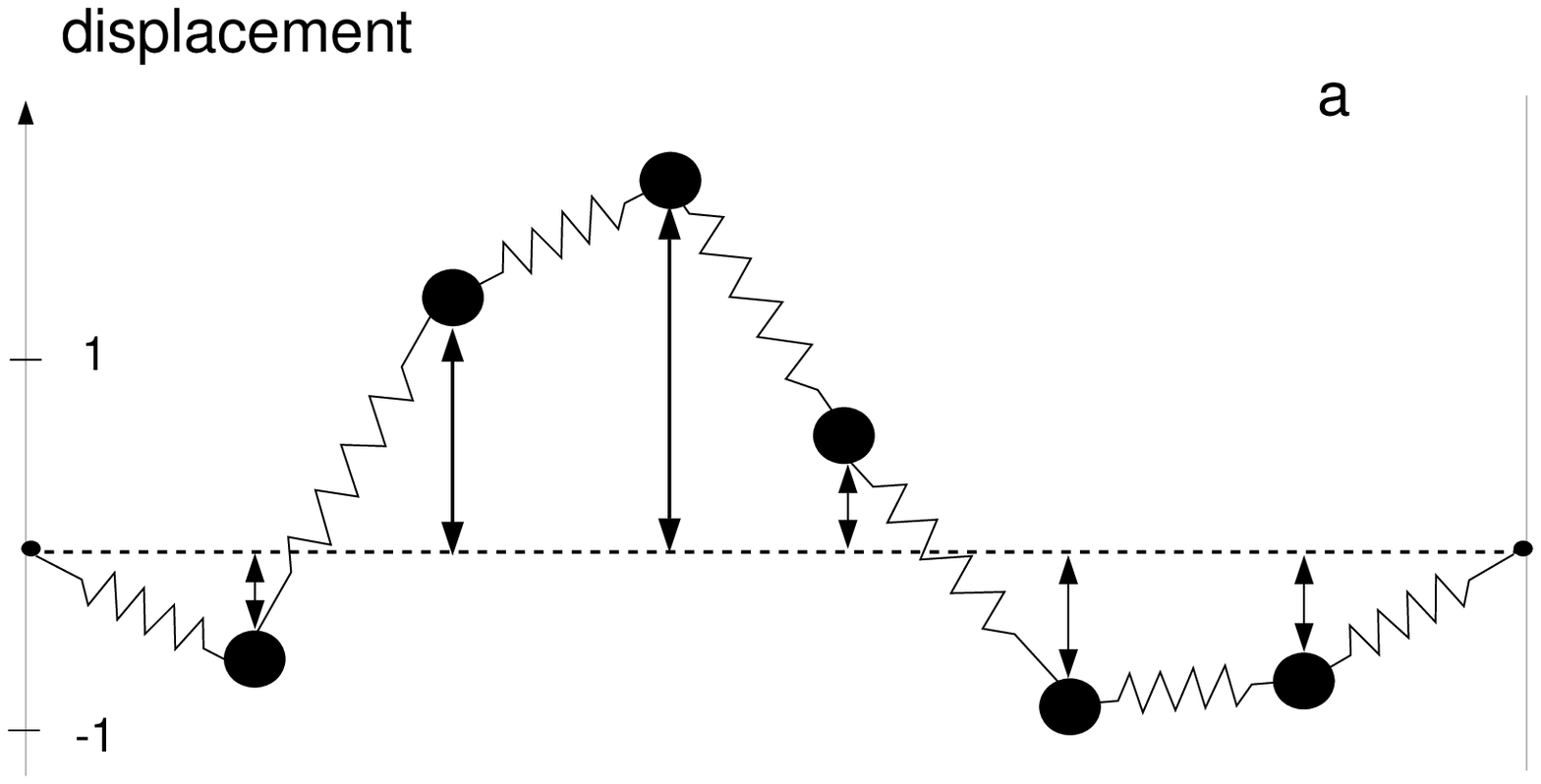, width=2.5in}
\hspace*{\fill}
\epsfig{file=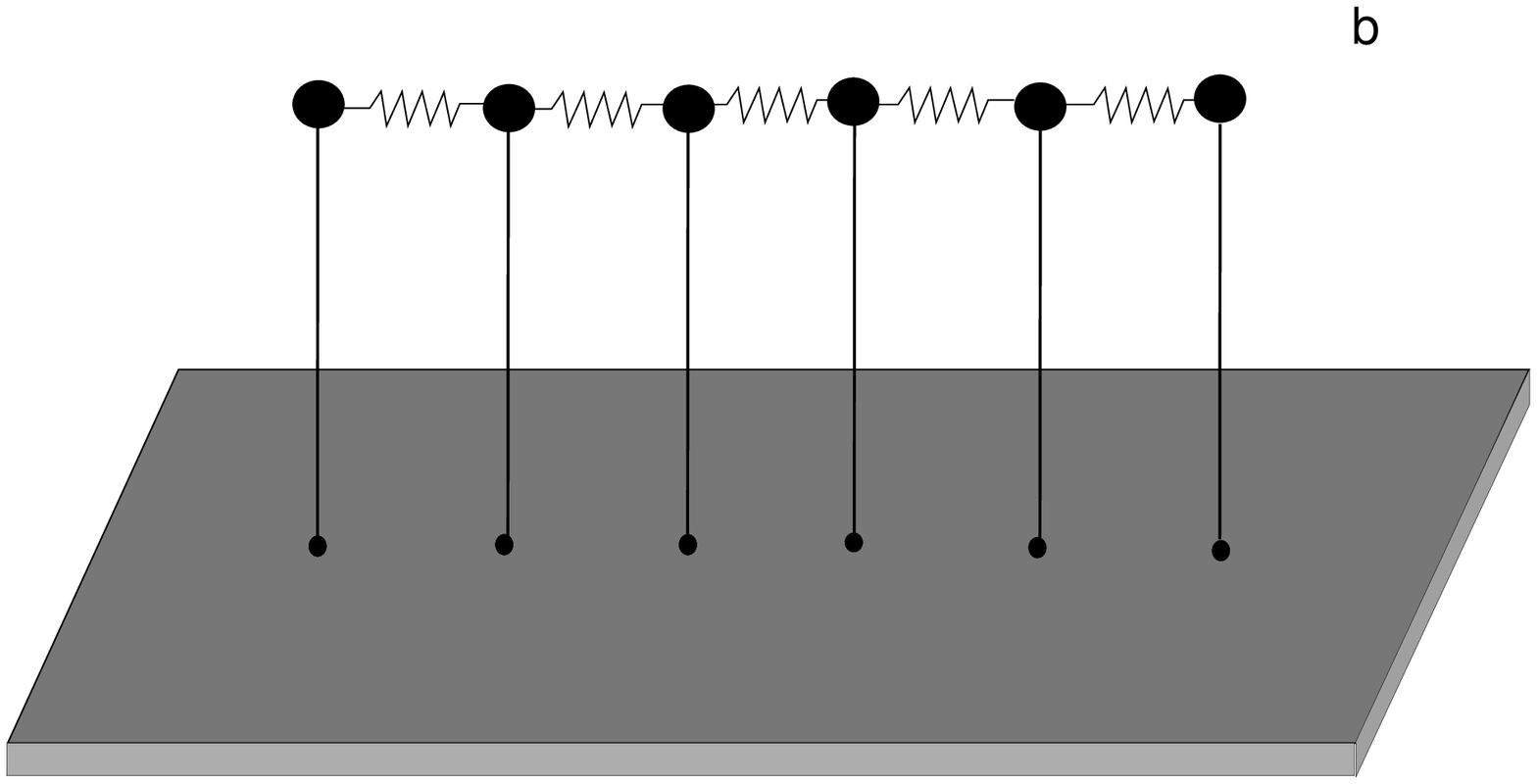, width=2.5in}
\hspace*{\fill}
\caption{\label{x:pendulum}\small An unstable dynamical system. 
a) The unstable chain with the mass points displaced from the unstable 
fixed point which is indicated by the horizontal dashed line. 
The masses are coupled to their neighbors with springs, and those 
at the end of the chain are connected to a rigid wall. 
b) A chain of upward-pointing pendulae connected by springs as an 
example of an unstable spatially extended system.}
\end{figure}

The system we studied, illustrated in Fig.~\ref{x:pendulum}a, consists 
of {\it n} mass
points connected to their neighbors by springs. In addition a
destabilizing force proportional to the displacement acts on each mass
point. This force models the behavior of unstable fixed points: the
force is zero exactly at the fixed point, but acts to amplify any small
deviations away from the fixed point. This system can be construed as a
linear approximation to the behavior of a variety of dynamical systems
near an unstable fixed point, such as the inverted pendulae shown in the
Fig.~\ref{x:pendulum}b. In the absence of
control, any small initial displacement away from the vertical position
rapidly leads to all the masses falling over. In this case, the lowest
mode consists of all the pendulae falling over in the same direction and
is the most rapidly unstable mode of behavior for this system. By
contrast, higher modes, operating at shorter length scales, consist of
the masses falling in different directions so that springs between them
act to reduce the rate of falling.

The system{'}s physical behavior is described by
\begin{enumerate}
\item the number of mass points {\it n}
\item the spring constant {\it k} of the springs
\item a destabilizing force coefficient {\it f}
\item a damping force coefficient {\it g}
\end{enumerate} 
We also suppose the mass of each point is equal to one. The resulting
dynamics of the unstable chain is given by\footnote{We used a
standard ordinary-differential-equation solver~\cite{shampine75} to 
determine the controlled system{'}s
behaviors.}~\cite{goldstein80}:
\begin{equation}\label{x:motion}\vcenter{\halign{\strut\hfil#\hfil&#\hfil\cr 
$\displaystyle{{{dx_{i}}\over{dt}}}$\hfilneg&$\displaystyle{{}=
v_{i}\hskip 0.265em }$\cr 
$\displaystyle{{{dv_{i}}\over{dt}}}$\hfilneg&$\displaystyle{{}=
k{\left( x_{i-1}\hskip -0.167em -\hskip -0.167em 
x_{i}\right) }+k{\left( x_{i+1}
\hskip -0.167em -\hskip -0.167em x_{i}\right) }
+fx_{i}-gv_{i}+H_{i}}$\cr 
}}\end{equation}
where \(
x_{i}\) is the displacement of mass point
{\it i}, \(
v_{i}\) is the corresponding velocity, and \(
x_{0}=x_{n+1}=0\) is the boundary condition. The \(
H_{i}\) term in Eq.~(\ref{x:motion})
is the additional control force produced by the actuator attached to
mass point {\it i}. We suppose the
magnitude of this control force is proportional to the power
\(
P_{i}\) used by the actuator. For reasons of simplicity we
use a proportionality factor of 1.

For these systems, the long time response to any initial condition
is determined by the eigenvalues of the matrix corresponding to the
right hand side of Eq.~(\ref{x:motion}).
Specifically, if the control force makes all eigenvalues have negative
real parts, the system is stable~\cite{hogg96b}.
The corresponding eigenvectors are the system{'}s modes. Thus to evaluate
stability for {\em all} initial conditions, we can use
any single initial condition that includes contributions from all modes.
If there are any unstable modes, the displacements will then grow. We
used this technique to evaluate stability in the experiments described
below.

\section{ A Power Market for Control}
The control problem is  how hard to push on the various mass points
to maintain them at the unstable fixed point. This problem can involve
various goals, such as maintaining stability in spite of perturbations
typically delivered by the system{'}s environment, using only weak control
forces so the actuators are easy and cheap to fabricate, continuing to
operate even with sensor noise and actuator failures, and being simple
to program, e.g., by not requiring a detailed physical model of the
system.

Computational markets are one approach to this control
problem~\cite{clearwater96,Ferguson88,huberman95b,Kurose89,Malone88,sutherland68,
waldspurger92,wellman93}.
As in economics, the use of prices provides a flexible mechanism for
allocating resources, with relatively low information
requirements~\cite{hayek78}: a single price
summarizes the current demand for each resource.

In designing a market of computational agents, a key issue is to
identify the consumers and producers of the goods to be traded. Various
preferences and constraints are introduced through the definition of the
agents{'} utilities. This ability to explicitly program utility functions
is an important difference from the situation with human markets.
Finally, the market mechanism for matching buyers and sellers must be
specified.

In the market control of smart matter treated here, actuators, or
the corresponding mass points to which they are attached, are treated as
consumers. The external power sources are the producers and as such are
separate from consumers. All consumers start with a specified amount of
money. All the profit that the producers get from selling power to
consumers is equally redistributed to the consumers. This funding policy
implies that the total amount of money in the system will stay
constant.

In the spirit of the smart matter regime, where control
computations are fast compared to the relevant mechanical time scales,
we assume a market mechanism that rapidly finds the equilibrium point
where overall supply and demand are equal. Possible mechanisms include a
centralized auction or decentralized bilateral trades or arbitrage. This
equilibrium determines the price and the amount of power traded. Each
actuator gets the amount of power that it offers to buy for the
equilibrium price and uses this power to push the unstable chain.

The utility function for using power
{\it P} reflects a trade-off between using
power to act against a displacement and the loss of wealth involved.
While a variety of utility functions are possible, a particularly simple
one for agent {\it i}, expressed in terms
of the price of the power, {\it p}, and the
agent{'}s wealth, \(
w_{i}\), is:
\begin{equation}\vcenter{\halign{\strut\hfil#\hfil&#\hfil\cr 
$\displaystyle{U_{i}=-{{1}\over{2w_{i}}}
pP^{2}+bP{\left| X_{i}\right| }
}$\cr 
}}\end{equation}
where
\begin{equation}\label{x:structure}\vcenter{\halign{\strut\hfil#\hfil&#\hfil\cr 
$\displaystyle{X_{i}=\sum _{j=1}^{n}
a_{ij}x_{j}}$\cr 
}}\end{equation}
is a linear combination of the displacements of all mass points that
provides information about the chain{'}s state. The parameter
{\it b} determines the relative importance
to an agent of responding to displacements compared to conserving its
wealth for future use.

Actuator {\it i} always pushes in the
opposite direction of \(
X_{i}\), i.e., it acts to reduce the value of
\(
X_{i}\). In this paper we focus on the simple case of purely
local control where \(
a_{ij}=1\) when \(
i=j\) and is 0 otherwise. Thus, consumer
{\it i} considers only its own displacement
\(
x_{i}\). For simplicity, we use an ideal competitive market
in which each consumer and producer acts as though its individual choice
has no affect on the overall price, and agents do not account for the
redistribution of profits via the funding policy. Thus a consumer{'}s
demand function is obtained by maximizing its utility function as a
function of power:
\begin{equation}\label{x:demand}\vcenter{\halign{\strut\hfil#\hfil&#\hfil\cr 
$\displaystyle{{{dU_{i}}\over{dP}}}$\hfilneg&$\displaystyle{{}=
-p{{P}\over{w_{i}}}+b{\left| X_{%
i}\right| }=0\hskip 0.212em {\ifmmode\Rightarrow\else$\Rightarrow$\fi}\hskip 0.212em 
P_{i}{\left( p\right) }=b{\left| 
X_{i}\right| }{{w_{i}}\over{%
p}}}$\cr 
}}\end{equation}
This demand function causes the agent to demand more power when the
displacement it tries to control is large. It also reflects the
trade-off in maintaining wealth: demand decreases with increasing price
and when agents have little wealth. The overall demand function for the
system is just the sum of these individual demands, giving
\begin{equation}\vcenter{\halign{\strut\hfil#\hfil&#\hfil\cr 
$\displaystyle{P^{{\rm demand}}{\left( p\right) }
={{b}\over{p}}\sum {\left| 
X_{i}\right| }w_{i}}$\cr 
}}\end{equation}

Similarly, each producer tries to maximize its profit
\(
\rho \) given by the difference between its revenue from
selling power and its production cost \(
C{\left( P\right) }\): \(
\rho =pP-C{\left( P\right) }\). To provide a constraint on the system to minimize
the power use, we select a cost function for which the cost per unit of
power, \(
C{\left( P\right) }/P\) increases with the amount of power. A simple example
of such a cost function is
\begin{equation}\vcenter{\halign{\strut\hfil#\hfil&#\hfil\cr 
$\displaystyle{C{\left( P\right) }={{1}\over{2
a}}P^{2}}$\cr 
}}\end{equation}
The parameter {\it a} reflects the relative
importance of conserving power and maintaining stability. We obtain the
producer{'}s supply function by maximizing its profit:
\begin{equation}\vcenter{\halign{\strut\hfil#\hfil&#\hfil\cr 
$\displaystyle{{{d\rho }\over{dP}}=p-{{dC}\over{%
dP}}=0\hskip 0.212em {\ifmmode\Rightarrow\else$\Rightarrow$\fi}\hskip 0.212em P{\left( 
p\right) }=ap}$\cr 
}}\end{equation}
This is the same for all producers, so the overall supply function is
then just
\begin{equation}\vcenter{\halign{\strut\hfil#\hfil&#\hfil\cr 
$\displaystyle{P^{{\rm supply}}{\left( p\right) }
=nap}$\cr 
}}\end{equation}

From this the price and amount of traded power is determined by the
point where the overall supply and demand curves intersect, i.e.,
\(
P^{{\rm demand}}{\left( p\right) }
=P^{{\rm supply}}{\left( p\right) }
\). For our choices of the utility and cost functions,
this condition can be solved analytically to give
\begin{equation}\vcenter{\halign{\strut\hfil#\hfil&#\hfil\cr 
$\displaystyle{p_{trade}=\Rad{{{b}\over{na}}
\sum _{i=1}^{n}{\left| X_{%
i}\right| }w_{i}}\DoRad }$\cr 
}}\end{equation}
Given this equilibrium price, agent {\it i}
then gets an amount of power equal to \(
P_{i}{\left( p_{trade}\right) }
\) according to Eq.~(\ref{x:demand}) and the resulting control force is directly proportional
to received power.

We can also consider the case where the amount of power available
to the system is limited to \(
P^{{\rm global}}_{{\rm max}}\). This hard constraint has the effect of limiting the
overall supply function when the price is high so it becomes
\begin{equation}\vcenter{\halign{\strut\hfil#\hfil&#\hfil\cr 
$\displaystyle{P^{{\rm supply}}{\left( p\right) }
=
\left\{\matrix{nap&{\rm if}\hskip 0.212em p{\ifmmode<\else$<$\fi}P^{%
{\rm global}}_{{\rm max}}/na\cr 
P^{{\rm global}}_{{\rm max}}&
{\rm otherwise}\cr }\right.}$\cr 
}}\end{equation}

The final aspect of the market dynamics is how the wealth changes
with time. This is given by
\begin{equation}\vcenter{\halign{\strut\hfil#\hfil&#\hfil\cr 
$\displaystyle{{{dw_{i}}\over{dt}}}$\hfilneg&$\displaystyle{{}=
-pP_{i}{\left( p\right) }+{{1
}\over{n}}pP^{{\rm demand}}{\left( 
p\right) }}$\cr 
$\displaystyle{}$\hfilneg&$\displaystyle{{}=-b{\left| X_{i}\right| }
w_{i}+{{b}\over{n}}\sum _{%
j=1}^{n}{\left| X_{j}\right| }
w_{j}}$\cr 
}}\end{equation}
because we use the funding policy that all expenditures are returned
equally to the agents in the system.

\section{ Comparing with Local Controls}
As a simple comparison for the market behavior, we also study a
local control method. In this case, each actuator
{\it i} pushes with a strength that is
proportional to the displacement of its respective mass point, and
ignores the displacements of all other mass points. Specifically, the
local controls are simply given by \(
H_{i}=-c\hskip 0.212em x_{i}\).  Other control strategies~\cite{hogg96b} 
can try to estimate the amplitude of the
lowest modes and push only against these modes, since these are the ones
most important for stability.

For comparison with the market, we restrict ourselves to the case
where the amount of available power is limited. This is useful for
evaluating the ability of different control methods to maintain
stability using only weak forces. We distinguish two ways the power
could be limited for the local control. In the first, each actuator is
separately limited to use no more than \(
P_{{\rm max}}\) power (local control 1), which corresponds to a
situation where each actuator has a separate power source such as its
own battery. Any actuator that requests more power than this maximum has
its control force reduced to require only \(
P_{{\rm max}}\), i.e., \(
{\left| H_{i}\right| }=P_{{\rm max}
}\). The second local control allows available power to
be moved among the different actuators and is limited only in that all
actuators together cannot use more than \(
P^{{\rm global}}_{{\rm max}}=nP_{{\rm 
max}}\) power (local control 2). This overall limit is
implemented by comparing the total power requested according to the
local control, i.e., \(
P_{{\rm request}}=c\sum _{i}
{\left| x_{i}\right| }\) to the maximum available. If the requested amount
exceeds the maximum, each agent has its power reduced by the factor
\(
P^{{\rm global}}_{{\rm max}}/P_{%
{\rm request}}\) so that the overall used power equals the global
limit. The corresponding market has a total available power of
\(
P^{{\rm global}}_{{\rm max}}=nP_{{\rm 
max}}\).

\section{ Results}
We studied a chain composed of 27 mass points, all of them having
unit mass and connected by springs with a spring constant of value 1 and
damping coefficient 0.1. The destabilizing force coefficient is 0.2,
which is sufficient to make the system unstable when there is no control
force. All agents start with an initial wealth of 50 money units and we
are using the values \(
a=0.05\) and \(
b=0.001\) in the cost and utility functions. For definiteness,
we chose an initial condition where the single element in the middle of
the chain had a unit displacement and all other values were at zero.
This configuration includes a contribution from all the modes of the
system, which are just sinusoidal waves in this chain with uniform
masses and spring constants~\cite{goldstein80}.
For the local control, we used \(
c=0.2\), which is more than sufficient to ensure stable
control when power is unlimited~\cite{hogg96b}.

\begin{figure}
\hspace*{\fill}
\epsfig{file=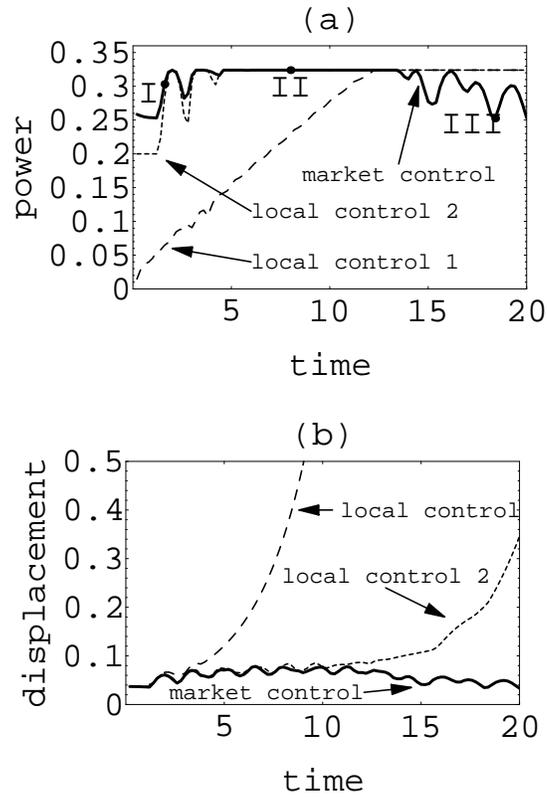, width=3in}
\hspace*{\fill}
\caption{\label{x:limit}\small a) Time development of the overall power usage 
for a market control (solid) and local controls 1 and 2 (dashed) in the 
case of limited available power. With the same power limit in all three 
cases, the market is the only one that can control the unstable chain. 
Points I, II and III mark the times at which the supply and demand curves 
intersections are shown in Fig.~\ref{x:curves}. b) Corresponding time 
development of the average displacement for a market control and local 
controls 1 and 2. The market reduces the average displacement with time 
whereas the local control is not able to prevent it from growing.}
\end{figure}

With \(
P_{{\rm max}}=0.012\), Fig.~\ref{x:limit} compares
the performances of both local and market controls. We show both the
total power use \(
\sum _{i}P_{i}\) and the average displacement of the chain
\(
\sum _{i}{\left| x_{i}\right| }
/n\). As can be seen, for the chosen parameter values the
market is able to control the unstable chain in spite of the fact that
the power is limited to a global maximum. This limit is reached several
times. The local controls (1 and 2), on the other hand, fail in both
cases, as seen in the figure. These results were obtained in a
simulation run that lasted 20 time units. A longer simulation shows that
the overall power usage and average displacements decrease with time for
the market control while displacement continues increasing for the local
controls.

\begin{figure}
\hspace*{\fill}
\epsfig{file=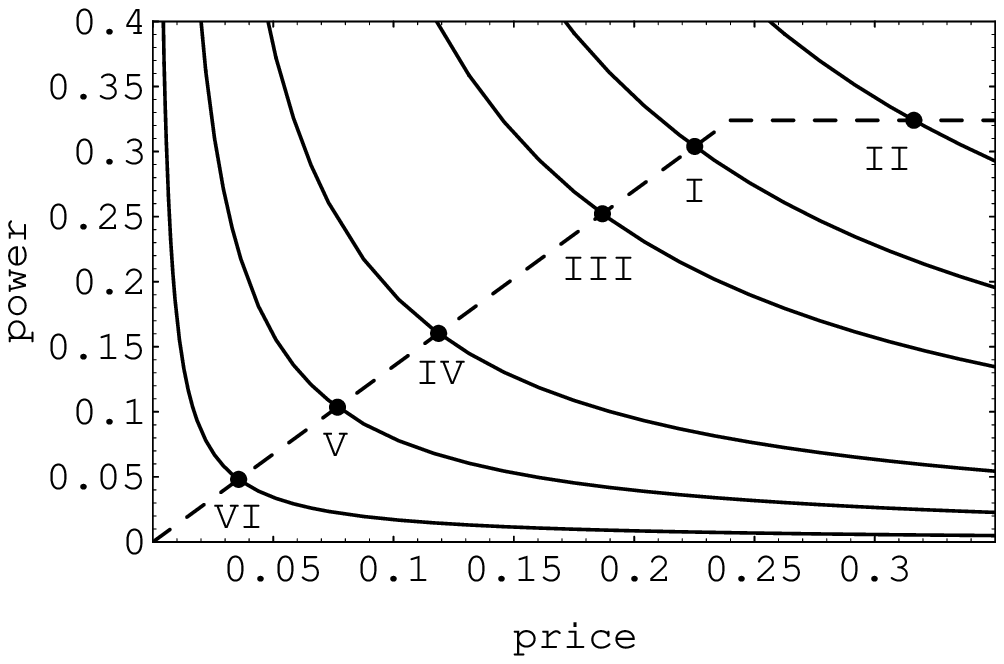, width=4in}
\hspace*{\fill}
\caption{\label{x:curves}\small Overall supply curve (dashed) and the overall 
demand curves (solid) at times I=1.6, II=8.0, III=18.4, IV=40.0, 
V=60.0 and VI=90.0 for the market example of Fig.~\ref{x:limit}.}
\end{figure}

Since the power cost function \(
C{\left( P\right) }\) does not change, the overall supply curve never
changes, as shown in Fig.~\ref{x:curves} which
displays the supply curve and some demand curves for different times.
The demand curves depend on the displacements and wealth of the agents.
Since these are dynamical variables, overall demand curve changes in
time. In addition to the times I, II and III marked in
Fig.~\ref{x:limit}a, we also plot the overall
demand curves for later times IV, V and VI. This shows that the amount
of traded power decreases with time while the unstable chain is
controlled by the market.

\begin{figure}
\hspace*{\fill}
\epsfig{file=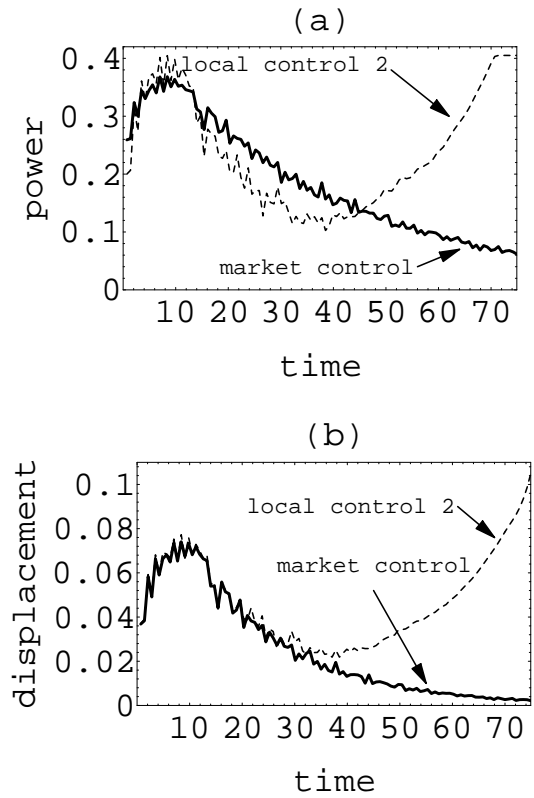, width=3in}
\hspace*{\fill}
\caption{\label{x:failure}\small Comparison of a market control and a local 
control in the case where one actuator breaks after 10 time units. 
Both control strategies would be able to control the system when all 
actuators would work perfectly. The market is still able to control 
the system although one actuator is broken but the local controller fails. 
a) Overall used power vs.~time. b)  Average displacement vs.~time.}
\end{figure}

To demonstrate how robust the market mechanism is, we show in
Fig.~\ref{x:failure} the system{'}s response when
an actuator breaks down. In this case we slightly increase the amount of
power available compared to the simulation used for
Fig.~\ref{x:limit} to \(
P_{{\rm max}}=0.015\), so that the local controller 2 can also control the
unbroken system. With the system initially functioning properly, we
turned off the actuator in the middle of the chain after 10 time units
and observed the consequent evolution. As can be seen, the market is
still able to control the system whereas the local control fails to do
so.

\section{ Discussion}
In this paper we presented a novel mechanism of controlling
unstable dynamical systems by means of a multiagent system using a
market mechanism. We described how we defined consumers{'} and producers{'}
utility functions that lead to the overall supply and demand curves and
evaluated the price and amount of traded power within the system.

We showed that the market approach is able to control an unstable
dynamical system in the case of limited power whereas a traditional
local control strategy fails under the same assumptions. We also
demonstrated that a market control adapts better to cases when an
actuator breaks during the controlling process. These results show that
a market control can be more robust than a local control when operating
with given power constraints by focussing the power in those parts of
the system where it is most needed. This not only reduces total power
use but, more importantly, also allows control with weaker, and thus
easier to fabricate, actuators.

The power of market approaches to control lies in the fact that
relatively little knowledge of the system to be controlled is needed.
This is in stark contrast to traditional AI approaches, which use
symbolic reasoning with extremely detailed models of the physical
system. However, while providing a very robust and simple design
methodology, a market approach suffers from the lack of a high level
explanation for its global behavior. An interesting open issue is to
combine this approach with the more traditional AI one.

Although we have chosen particular forms of utility, supply and
demand functions, there are many other functional forms that can also
control the system. These could include additional goals, such as faster
recovery from sudden changes and minimizing the number of active
actuators. Furthermore, different funding strategies are possible, where
profits are shared unequally among agents or the funds are allocated by
an external agent. A very promising approach is the possibility of
improving the performance of the system by having different market
organizations that change in time. In our system, this corresponds to
the agents learning to use information on the displacements or
velocities of their neighbors when making their control decisions. In
this way the multiagent system would take advantage of the fact that
markets are a simple and powerful discovery process: new methods for
selecting trades can be tried by a few consumers or producers and, if
successful relative to existing approaches, gradually spread to other
agents. Such a learning mechanism could help the system discover those
organizational structures that lead to improved performance and
adaptability.

\end{document}